# Physical probes expose and alleviate chemical-environment collapse in molecular representations

Jiebin Fang,[1,2] Zidi Yan,[1,2] Churu Mao,[2] Yongjun Jiang,[3] Xinyi Tang,[2] Lei Miao,[4] Dan Lu,[5] Yun Huang,[2*] Wanjing Ding[1,2]* & Zhongjun Ma[1,2]*

[1] Hainan Institute, Zhejiang University, Sanya 572025, China
[2] Institute of Marine Biology and Pharmacology, Ocean College, Zhejiang University, Zhoushan 316021, China
[3] School of Food and Pharmacy, Zhejiang Ocean University, Zhoushan 316021, China
[4] Hangzhou Bio-Sincerity Pharmaceutical Technology Company Limited, Hangzhou 311103, China
[5] School of Pharmaceutical Sciences, Shanghai Jiao Tong University, Shanghai 200240, China

[*]Email: h_yun@zju.edu.cn (Y. H.); mazj@zju.edu.cn (Z.M.); wading@zju.edu.cn (W. D.);

**ABSTRACT:** Nuclear magnetic resonance (NMR) spectroscopy provides an experimental readout of local chemical environments, but its use in molecular representation learning has been constrained by heterogeneous data and incomplete atom-level assignments. Here we construct complementary high-fidelity experimental and computational $^{13}$C NMR resources, which reveal a recurrent form of representational collapse: atoms that are equivalent in molecular topology can remain experimentally distinct in their real chemical environments, whereas explicit 3D descriptions are further limited by static conformations in dynamic regimes. To alleviate this bottleneck, we develop CLAIM (Contrastive Learning for Atom-to-molecule Inference of Molecular NMR), a framework that aligns efficient topological molecular inputs with atom-resolved NMR observables. Through hierarchical chemical priors and cross-level contrastive learning, CLAIM restores lost chemical resolution and markedly improves atom-level molecule–spectrum retrieval. CLAIM remains robust in flexible and tautomeric systems for $^{13}$C NMR prediction, improves stereoisomer discrimination without explicit 3D modelling, and transfers to broader molecular property tasks including ADMET prediction and fluorescence estimation. These results establish physically grounded spectral alignment as an effective strategy for alleviating chemical-environment collapse and for guiding experimentally grounded molecular representation learning.

## Introduction

In the wave of AI for Science, constructing models capable of precisely mapping microscopic chemical space to macroscopic physicochemical properties remains a core vision of computational chemistry[1, 2]. Whether for the high-throughput screening of biological activity in drug design[3] or the precise calculation of Stokes Shift in optoelectronic material development[4], the essence relies on the accurate prediction of molecular properties. A quintessential scientific challenge in these optimization processes lies in navigating the delicate balance of chemical modifications: efficiently enhancing target biological activity through minute structural perturbations while rigorously preserving established macroscopic physicochemical profiles[5]. To successfully decouple and predict these complex, often divergent structure-property relationships, computational models must possess exceptionally high structural sensitivity and resolution[6]. While deep learning has made significant progress in structure-activity relationships, molecular property prediction, and structural design in recent years[6, 7, 8, 9, 10], obtaining molecular representations with higher chemical resolution remains a critical step toward overcoming current data bottlenecks and improving prediction accuracy within the data-driven paradigm[11, 12].

Reviewing the development of molecular representation technology, we find that it is not simply moving toward higher dimensions but has instead become constrained by a specificity–dynamics trade-off. On one hand, topological representations, including molecular fingerprints (e.g., ECFP[13]), SMILES[14 15], and molecular graphs[16, 17], leverage rapid encoding of local substructures while implicitly encompassing the full conformational space accessible to a molecule. Owing to their high computational efficiency, they remain important tools in the industry today[18, 19]. However, this universality comes at the cost of resolution, termed topological degeneracy, leading to limited capability in representing fine-grained structural features such as stereochemistry[20]. On the other hand, to compensate for the lack of spatial information in topological descriptors, geometric deep learning introduces explicit 3D coordinates[21] or quantum features based on electron cloud density[22, 23]. Although this significantly improves conformational specificity[24], it falls into the static geometric trap: relying on a single static low-energy conformer fails to truthfully reflect the dynamic average behavior presented by molecules in solution due to rapid rotation or tautomerism[25]. A central challenge, therefore, is to obtain a latent representation that combines the stereochemical discriminability of geometric representations with the implicit robustness of topological representations toward dynamic ensembles.

This challenge has motivated growing interest in end-to-end self-supervised learning for molecular representation[11, 26]. In this context, contrastive learning has emerged as a powerful framework for unifying different data views in scientific computing[11]. Just as CLIP established the cornerstone of cross-modal representation in computer vision through image-text alignment[27], DrugCLIP[28] and CLOOME[29] have respectively validated the effectiveness of contrastive learning in mapping chemical structures to protein pockets or biological phenotype spaces. Recently, multimodal pre-training has further attempted to fuse explicit geometric conformers, pharmacophores, and topological graphs as different modalities, attempting to reconstruct multidimensional features of molecules in latent space[30, 31]. However, such methods remain essentially complementary integrations of different computational views[2]. To break the bottleneck of physical precision, a higher-order supervisory signal is required—one that is not merely another computational representation, but a direct physical mapping of molecular essence. NMR spectra, as experimentally grounded fingerprints of molecules, provide an attractive supervisory signal by offering atom-level resolution while capturing time-averaged chemical environments[32, 33]. This suggests that aligning SMILES-based molecular representations with NMR spectra—physical observables that reflect dynamic conformational ensembles and electronic environments—may provide an effective route toward higher-precision molecular representation.

Despite these opportunities, the application of contrastive learning to NMR remains constrained by severe data heterogeneity. Existing efforts such as NMRextractor[34] and NMRexp[35] have utilized automated pipelines to build million-level experimental record libraries; however, these automatically collected datasets typically lack atom-level assignment between spectral signals and molecular atoms. Studies in multimodal representation learning have shown that noisy correspondence can disrupt alignment structure in the feature space, leading to substantial losses in generalization for fine-grained tasks[36]. This limitation is particularly evident in recent NMR applications: while pioneering studies like CRESS and NMRMind rely on coarse-grained full-spectrum matching[37, 38], frameworks like NMR-Solver achieve significantly better structural resolution precisely by explicitly modeling atom-level assignment[39]. Although DFT-generated data provide clear atom-wise correspondence[40], they also introduce a substantial distribution shift relative to experimental spectra, which leads to pronounced performance decay when models trained on computational data are transferred to real experimental settings [41, 42]. Thus, the combination of massive but unassigned experimental data and assignment-complete yet biased computational data creates a major training bottleneck for learning fine-grained, physically meaningful molecular representations.

To overcome this bottleneck, we first established two complementary $^{13}$C NMR resources that make the representational failure itself observable. Through hierarchical chemical priors and atom-resolved curation, we constructed NMRSDB-Exp, a high-fidelity experimental dataset with explicit atom-level correspondence. Importantly, this process did more than improve data quality: once noisy and misassigned records were resolved, a recurrent mismatch became apparent between topological equivalence and real chemical environments, as atoms indistinguishable in 2D connectivity often exhibited clearly separated experimental $^{13}$C shifts. In parallel, we developed the AutoStereoQ workflow to generate NMRSDB-Cal, a computational benchmark for controlled evaluation of stereochemical and conformational complexity. This benchmark further shows that although explicit 3D approaches can increase structural specificity, their ability to resolve such cases remains constrained by static approximations, dynamic averaging, and computational cost.

Built on these resources, we developed CLAIM (Contrastive Learning for Atom-to-molecule Inference of Molecular NMR), a framework that uses atom-resolved experimental spectra as physical supervision to align efficient topological inputs with real chemical environments. Rather than relying on explicit 3D calculations, CLAIM learns a latent space guided by the time-averaged molecular information encoded in 1D NMR observables, thereby disentangling chemically distinct environments that remain unresolved under topology-only descriptions. We evaluated this representation in three progressively broader settings: near-domain validation in $^{13}$C NMR prediction, where CLAIM shows particular robustness in flexible and tautomeric regimes; reverse inference in stereoisomer retrieval, where a delta-learning extension improves discrimination beyond topology-only representations; and forward transfer to broader molecular property tasks, including ADMET and fluorescence prediction. Together, these results support physically grounded spectral alignment as an effective strategy for alleviating chemical-environment collapse in molecular representations.

## Results and Discussion

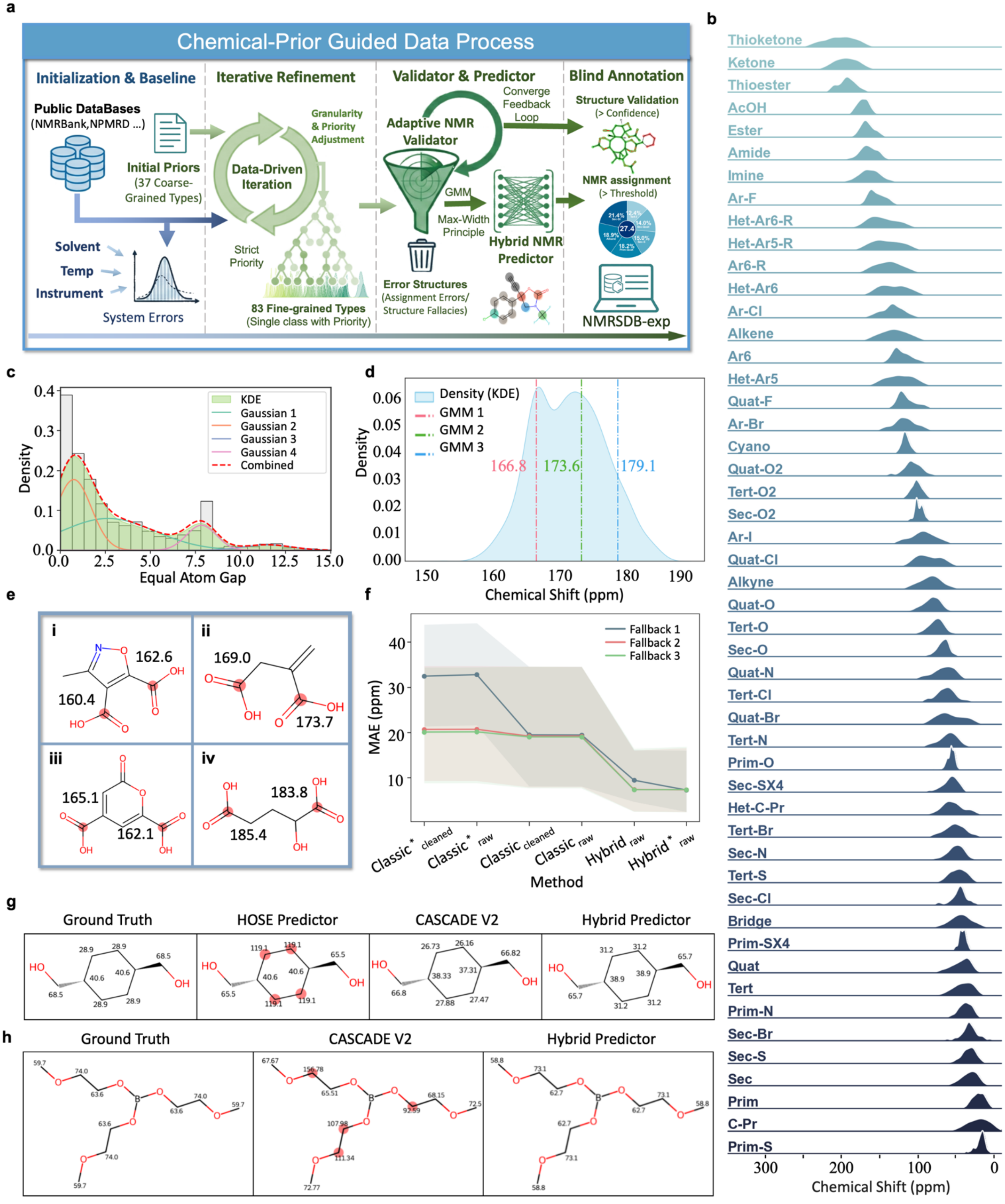

**Figure 1. Data cleaning workflow based on chemical priors and hybrid predictor for the construction of the NMRSDB-exp database. a.** Flowchart of chemical prior-guided data processing. The workflow first utilizes public databases[43] to establish baseline distributions for systematic errors such as instrument and solvent effects; subsequently, it refines fine-grained atom type chemical priors via a data-driven iterative optimization strategy, which are used to construct the Validator and Predictor; finally, the NMRSDB-exp database is generated through blind annotation and data cleaning of open-source databases[34, 44]. **b.** $^{13}$C NMR chemical shift distributions of the 50 most frequent atom types. Probability distributions were fitted using Kernel Density Estimation (KDE) based on cleaned NMRShiftDB2[43] data. **c.** Distribution of chemical shift differences for topologically equivalent but chemically inequivalent atoms. Chemical shift differences between topologically equivalent atoms were calculated via simulated labeling and random sampling of carbon atoms. Four main contributing components were identified using Gaussian Mixture Model (GMM) fitting. **d.** Heterogeneity of chemical environments and GMM fitting. Taking the carboxylic acid $\alpha$-carbon atom as an example, this panel illustrates the multimodal characteristics of the chemical shift distribution within a single atom type. The

blue curve represents the KDE of experimental data, while dashed lines show the three principal gaussian components obtained via GMM decomposition. **e.** Representative structural cases of carboxylic acid $\alpha$-carbon atoms. Four typical molecular structures corresponding to different gaussian component peaks within the range of panel **d** are displayed. **f.** Comparison of Mean Absolute Error (MAE) for NMR predictors. The panel compares performance under different training strategies (Classic/Hybrid) and datasets (Raw/Cleaned). The superscript "*" indicates that the training set includes test data (full training), and the shaded area represents the distribution range of absolute errors. *Fallback N* denotes the strictness of the lookup strategy: *Fallback 3* requires matching at least 3 high-precision records before returning a result, otherwise degrading the match; *Fallback 1* requires matching only 1 record. **g-h**. Comparison of NMR prediction results for a cyclohexane derivative (**g**) and a flexible molecule (**h**). For the cyclohexane structure in g, the prediction values from the HOSE predictor, CASCADE-2.0[45], and the Hybrid Predictor are compared with experimental ground truth, where the HOSE predictor exhibits significant errors. For the flexible molecule in h, CASCADE-2.0 shows substantial deviations and differentiates equivalent atoms. Red circles highlight atomic positions where the prediction gap exceeds 17 ppm.

## Establishing high-confidence NMR observations

We first sought to distinguish measurement variation from differences arising from genuine chemical environments (Fig. 1a), thereby establishing a reliable basis for representation learning from NMR spectra. Before the limitations of molecular representations can be assessed against experimental observables, experimental heterogeneity itself must be controlled. Systematic sources of variation, including solvent, temperature and instrumental effects, were estimated by analyzing the distribution of chemical-shift differences for atoms in identical chemical environments across diverse experimental records. The resulting distribution contained both systematic variation and non-Gaussian components associated with structure-dependent splitting in specific samples. A Gaussian mixture model (GMM) was therefore used to separate these effects (Fig. 1c), yielding a more stable estimate of the systematic-error baseline.

On this basis, we constructed an initial chemical-prior system for carbon environments. Starting from 37 coarse-grained categories, we used an iterative data-driven refinement procedure to split overly broad classes, resulting in 83 atom-environment categories together with a priority-based arbitration scheme to reduce matching ambiguity. Analysis of the cleaned NMRShiftDB2[43] dataset, comprising 226,969 carbon atoms (Fig. 1b), showed that most atom types occupied characteristic chemical-shift ranges consistent with established chemical expectations. Aliphatic carbons were concentrated in the upfield region, aromatic and olefinic carbons occupied intermediate ranges, and carbonyl-related groups remained predominantly downfield. These priors therefore provide chemically interpretable boundaries for evaluating whether observed shift differences are likely to reflect noise or genuine local nonequivalence.

Because unimodal normal envelopes were insufficient for high-precision anomaly detection, we further developed the Adaptive NMR Validator. This framework combines GMM-based modeling with a coverage-driven feedback loop based on kernel density estimation to define confidence boundaries for heterogeneous chemical environments. Using a stringent low-confidence threshold, the validator identified 2,169 atoms for retrospective inspection in public datasets. Subsequent analysis revealed recurrent inconsistencies, including atom-index misalignment and structural misannotation, indicating that data quality issues extend beyond simple measurement noise. Together, these priors and confidence estimates established the atom-level reliability required for subsequent curation, and thereby created the conditions under which a hidden mismatch between topology and chemical environment could become observable in experimental data.

## Atom-level assignment reveals chemical-environment collapse in topological representations

Although the Adaptive NMR Validator reduced explicit structural errors, it also made a recurrent representational bottleneck easier to detect: public repositories such as NMRexp[35], NPMRD[44] and NMRbank[34] generally lack the atom-to-peak assignments needed to determine when topologically equivalent atoms are experimentally nonequivalent. Because this representational bottleneck is expressed at the level of individual atoms, it cannot be assessed reliably without recovering atom-level correspondence from experimental spectra. Reconstructing these mappings by global bipartite matching is challenging when predictor uncertainty is large relative to the intrinsic shift range of a given atom type. In practice, conventional topology-based predictors[46] provided insufficient precision for reliable assignment in narrow-range environments. To improve assignment fidelity, we introduced a Hybrid NMR Predictor that combines topology-based prediction with the chemical priors defined above. This strategy constrains predicted values to chemically plausible intervals and reduces implausible matches arising from latent dataset noise or topological overgeneralization.

Importantly, lower regression error alone does not guarantee better performance in automated assignment. The curation pipeline relies on rank-based bipartite matching together with a predefined distinguishability threshold, such that uncertain candidate matches are discarded rather than forcibly assigned. In this setting, explicitly geometry-based models do not necessarily resolve the assignment bottleneck[47]. As shown in Fig. 1h, for flexible molecules these models can spuriously differentiate atoms that are chemically equivalent in solution because they are evaluated on rigid conformational snapshots. This introduces artificial shift gaps, compresses the effective discrimination interval, and can trigger incorrect ranking decisions that pass the matching criterion. By contrast, prior-guided topological methods preserve true chemical equivalence and avoid this false spatial differentiation, which makes them better suited to high-precision automated annotation despite their lower nominal regression accuracy.

Under these prior-guided constraints, assignment accuracy improved substantially. The Hybrid NMR Predictor reduced the overall MAE from 13.5 ppm to 4.3 ppm and decreased the MAE for outlier cases from 19.3 ppm to 8.3 ppm, while increasing generalization coverage on structurally novel molecules from 93.4% to 99.1%. Using this pipeline, we curated 25,364 reliable entries from NMRShiftDB2[43], annotated 11,808 high-confidence data points from NMRBank[34], and assigned 476 entries from NPMRD[44], yielding the NMRSDB-Exp dataset with 37,132 atom-level assigned experimental records.

Importantly, this curation process did not merely improve annotation quality; it repeatedly exposed a hidden representational failure that had been obscured by noisy and unassigned data. During atom-level assignment, we repeatedly encountered cases in which atoms that are indistinguishable in 2D topology exhibited clearly different $^{13}C$ shifts. This pattern was evident in the multimodal splitting of carboxylic acid $\alpha$-carbons (Fig. 1d,e) and recurred across five common motif classes, including allylic carbons, endocyclic and exocyclic carbons in symmetric rings, conformationally constrained $\alpha$-amino carbons, and vicinal carbons adjacent to chiral centres. Across these motifs, atom pairs that are equivalent in graph connectivity showed shift differences on the order of 4–11 ppm, consistent with stereoelectronic and conformational effects that are not captured by purely topological representations. These observations indicate that 2D topological equivalence can collapse chemically distinct environments, motivating the examination of explicit 3D representations as the most immediate alternative.

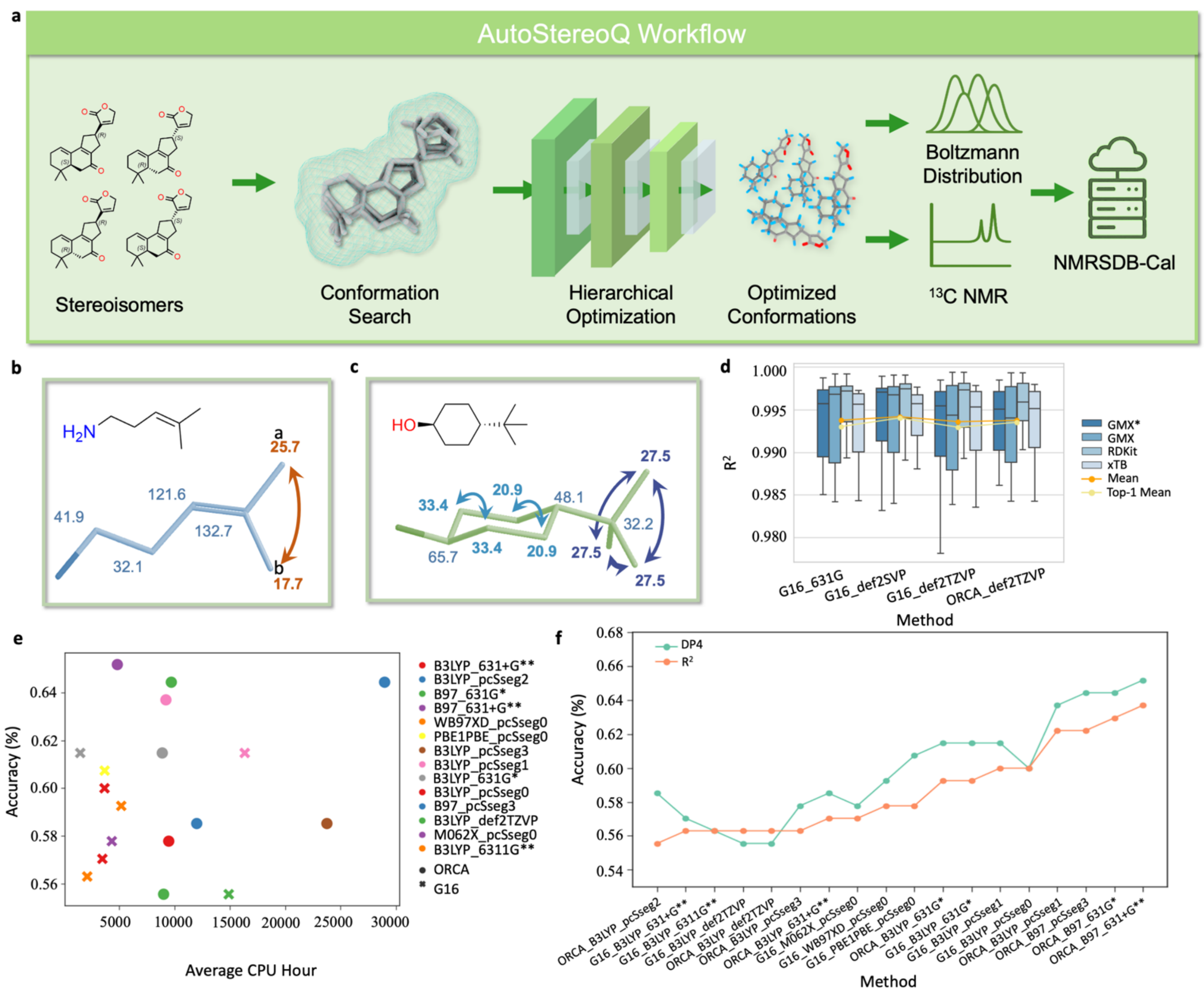

**Figure 2. AutoStereoQ workflow, construction of the computational NMR database (NMRSDB-cal), and benchmarking. a.** Schematic diagram of the fully automated AutoStereoQ processing workflow. It covers the pipeline from stereoisomer enumeration and verification, conformational search[48], and hierarchical structure optimization, to the final Boltzmann-weighted GIAO-NMR calculation. **b**. Topologically equivalent but chemically inequivalent: Two methyl groups on an alkene share identical graph topological connectivity but exhibit significantly different chemical shifts. **c**. Chemically equivalent: Symmetric carbon atoms on a cyclohexane skeleton and terminal tert-butyl groups exhibit completely consistent chemical shifts. **d.** Influence of geometric optimization precision on results. The orange line represents results from the weighted average of the conformational ensemble, while the yellow line represents results from only the thermodynamically optimal (Top-1) conformer. Ensemble averaging significantly improves prediction robustness. **e.** Cost–accuracy trade-off of DFT-based stereoisomer recognition methods. Each point represents one functional/basis-set/software combination, showing the relationship between average CPU time for the NMR calculation stage and retrieval accuracy. **f.** Benchmarking of DFT stereoisomer retrieval methods, including DP4[49] evaluation. Since enantiomers possess identical scalar NMR properties in achiral environments, the ground truth and its enantiomer were treated as the same class during evaluation.

### Explicit 3D modeling improves specificity but remains limited in dynamic regimes

Because the collapse exposed during atom-level curation arises from missing structural specificity in 2D topology, explicit 3D representations provide the most immediate route beyond topological degeneracy. In cases such as 4-methylpent-3-en-1-amine (Fig. 2b), atoms with identical 2D connectivity occupy distinct spatial and magnetic environments under stereochemical constraints, indicating that greater structural specificity is sometimes required to resolve NMR inequivalence. However, single static conformers are also insufficient, because they can introduce spurious geometric inequivalence by neglecting molecular motion. As illustrated in Fig. 2c, the methyl groups of (1R,4R)-4-(tert-butyl)cyclohexan-1-ol appear asymmetric in an optimized 3D snapshot, although rapid bond rotation renders them chemically equivalent in solution. These examples indicate that NMR-relevant environments are better described by conformational ensembles than by individual structures.

How such ensembles should be constructed was first evaluated by comparing four conformational search strategies across 11 structurally diverse natural products. Ensemble-averaged predictions consistently outperformed those based on the lowest-energy conformer alone (Fig. 2d), supporting the view that averaging across accessible states provides a more realistic description of solution-state NMR behavior. Systematic search strategies offered broader conformational coverage than purely trajectory-based approaches, whereas trajectory-based methods were more prone to local trapping. Together, these results support the use of ensemble-based explicit 3D representations when greater stereochemical specificity is required, while also indicating that static single-conformer descriptions are insufficient for the present problem.

We developed AutoStereoQ, an automated workflow spanning stereoisomer enumeration, geometry optimization, GIAO-NMR calculation and Boltzmann aggregation, to quantify the practical limits of this strategy at scale. This system was used to generate NMRSDB-Cal, a benchmark comprising 1,796 stereochemically complex molecules and 18 computational strategy combinations. Conformational optimization and screening were carried out over more than one year, and the subsequent NMR calculations on optimized conformers required 107,505 CPU hours. Across these settings, regression against experimental shifts remained uniformly high ($R^2 > 0.98$), while the relationship between computational cost and stereoisomer retrieval accuracy indicated only limited gains from increasingly expensive strategies (Fig. 2e). In stereoisomer retrieval, explicit DFT-based workflows improved performance over random guessing, but gains saturated rapidly: even with DP4[49, 50], accuracy reached a practical ceiling of 65.2% (Fig. 2f). Thus, although explicit 3D modeling improves structural specificity, increasing geometric fidelity alone does not fully resolve the chemical-environment collapse exposed by experimental NMR, especially in dynamic settings where solution-state averaging becomes important.

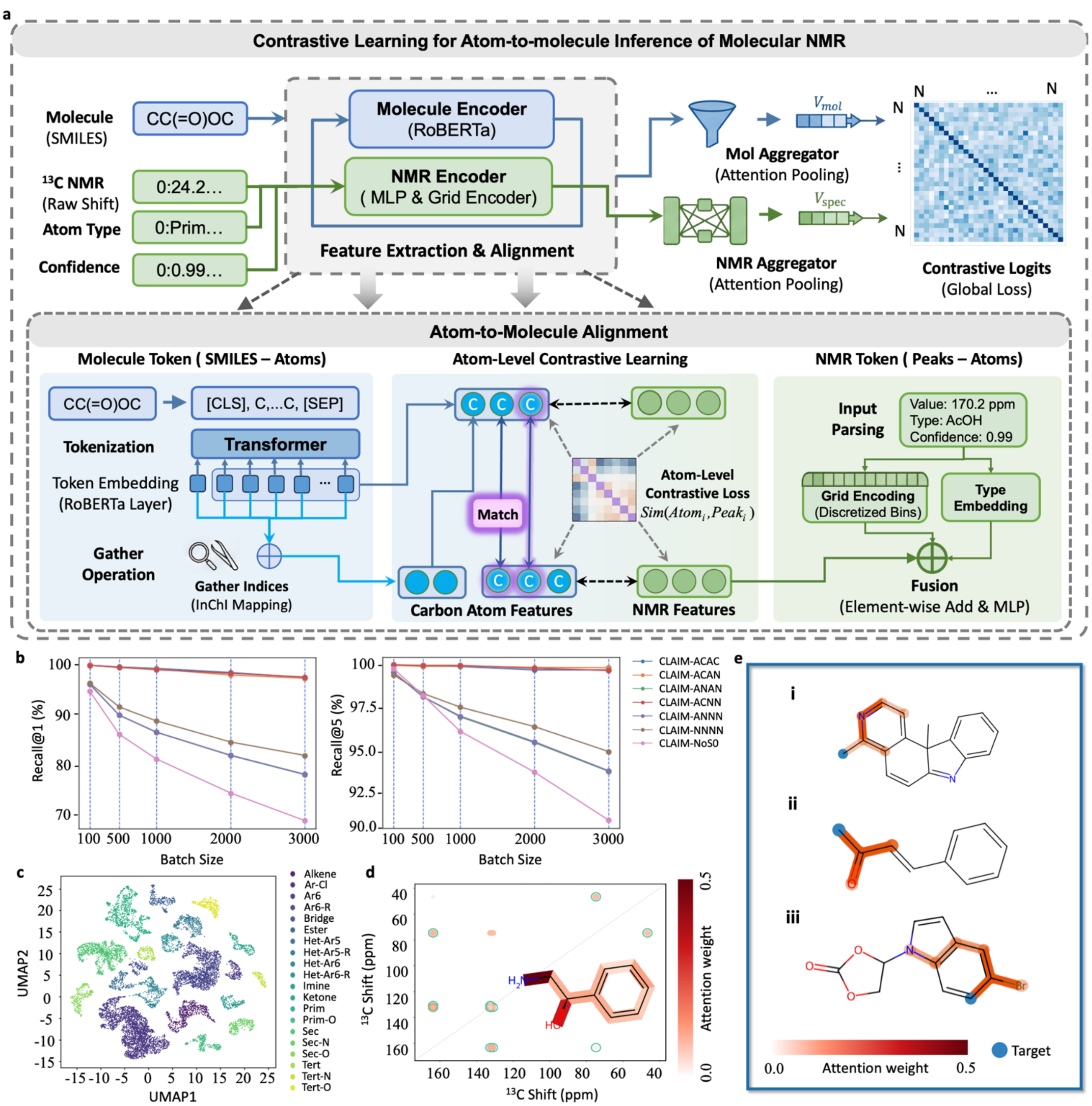

**Figure 3. Schematic diagram of the CLAIM framework architecture, retrieval performance evaluation, and interpretability visualization. a.** Architecture of the CLAIM model. The model consists of a dual-tower structure composed of a molecular encoder and an NMR encoder. The flowchart shows the data flow from SMILES sequence and NMR data inputs ($^{13}$C chemical shifts, atom types, and confidence scores) to two training stages (Stage 0: atom-level alignment and Stage 1: Global Contrastive Learning), and finally the process of generating molecule-level representations ($V_{mol}$ and $V_{spec}$) *via* attention pooling[51]. **b.** Recall@1 and Recall@5 retrieval accuracy of the CLAIM model and its variants under different batch sizes. Variant names encode their feature configurations using a four-letter suffix. The first and second letters indicate whether atom type (A) and confidence (C) are utilized or masked (N) during the Stage 0 atom-level alignment. The third and fourth letters indicate the status of these features during the Stage 1 molecule-level alignment. CLAIM-NoS0 is a baseline omitting Stage 0 entirely. Results are based on a test set of 3,714 independent molecule–spectrum pairs. **c.** UMAP[52] dimensionality reduction scatter plot of atom-level feature embeddings. Displays the clustering distribution of different atom type features in the latent space. **d.** Scatter plot correlating carbon atom pair attention weights with chemical shifts. Red spot intensity corresponds to the magnitude of attention weights, and green circles mark actual topological chemical bonds. The inset illustrates the mapping of attention weights onto the molecular structure, with bond colors ranging from light to deep red indicating weights from low to high. **e.** Atomic attention weight heatmaps within the SMILES encoder. Blue dots indicate target atoms, and the heatmap intensity corresponds to the attention weight distribution of other atoms in the molecule when constructing the feature of the target atom.

## CLAIM restores chemical resolution through physical alignment

To address the computational cost and error propagation associated with explicit 3D modeling, we developed CLAIM (Contrastive Learning for Atom-to-molecule Inference of Molecular NMR). Rather than relying on explicit 3D coordinates, CLAIM adopts a data-driven, top-down approach: it uses experimental NMR spectra as physical supervision to learn representations that align topological inputs with NMR-observable chemical environments and disentangle states that remain degenerate under topology-only encoding.

CLAIM adopts an asymmetric dual-tower architecture, with a molecular encoder for SMILES and an NMR encoder for spectral data, trained by atom-level (Stage 0) and molecule-level (Stage 1) contrastive learning (Fig. 3a). Ablation results show that Stage 0 is essential for retrieval precision. When atom-level alignment is removed and only Stage 1 is retained, recall@1 drops from 99.23% to 81.17% at Batch Size 1000, and from 97.40% to 68.83% at Batch Size 3000. By contrast, recall@5 decreases less markedly, from 99.83% to 93.70% at Batch Size 1000. This pattern indicates that molecule-level supervision preserves coarse matching, but atom-level alignment is important for disentangling hard negatives that remain unresolved under topology-only descriptions and for achieving reliable rank-1 retrieval.

On the spectral side, CLAIM separates the two components of NMR input: chemical-shift values and atom-level assignments. Continuous shifts are discretized on a high-resolution grid, while atom-environment types are encoded explicitly as semantic anchors. This design is motivated by the fact that atom-level assignments are experimentally accessible from multidimensional NMR measurements such as HSQC and HMBC and are routinely used in structure elucidation. The ablation results support their role in the model. Removing atom-type information reduces recall@1 from 99.23% to 86.50% at Batch Size 1000 and from 97.40% to 78.03% at Batch Size 3000, whereas the corresponding drop in recall@5 is smaller (99.83% to 96.93% at Batch Size 1000). Thus, atom-type information mainly improves rank-1 discrimination among locally similar chemical environments, helping the model disentangle states that remain topologically degenerate at the graph level.

We further introduced confidence scores from the Adaptive NMR Validator as dynamic training weights to account for quality variation in experimental data. Their effect is more limited than that of atom types and depends on the stage at which they are provided. When atom types are retained, adding confidence yields little additional benefit: ACAC and ACAN remain similar across settings, with recall@1 differing by 0.24% at Batch Size 1000 (99.23% versus 98.99%) and by 0.43% at Batch Size 3000 (97.40% versus 96.97%). By contrast, the gap between ACAC and ANAN is much larger, reaching 12.73% at Batch Size 1000 and 19.37% at Batch Size 3000 in recall@1. These results indicate that confidence contributes less than atom-level assignment within Stage 0. Its main contribution is to improve robustness under more difficult retrieval conditions, whereas the primary gain in chemical resolution arises from atom-level supervision itself, enabling CLAIM to disentangle chemically distinct environments that are otherwise merged under conventional topological representations.

**Physically aligned representations capture chemically meaningful structure**

We then conducted a multi-level interpretability analysis to determine whether CLAIM disentangles chemically meaningful local environments rather than merely fitting numerical correlations. At the local feature level, a UMAP[52] projection of atom-level feature embeddings (Fig. 3c) reveals the organization of the learned latent space: chemical semantics are effectively encoded, such that atoms in similar chemical environments—across different molecules—cluster spontaneously, forming a clear and continuous manifold. This organization suggests that the learned representation reflects shared local chemical environments across molecules while disentangling states that would otherwise remain degenerate under topology-only encoding.

Figure 3d illustrates the model's ability to recover chemically meaningful connectivity patterns. The analysis indicates that while primary amines and hydroxyl groups shift the chemical signal of their attached carbons downfield *via* inductive effects, the model also captures the deshielding effect produced by the magnetic anisotropy of the benzene ring on adjacent exocyclic structures. This trend suggests that the attention distribution is modulated by chemically relevant electronic effects rather than distance alone, consistent with the disentangling of environment-sensitive distinctions beyond pure topology.

At the global contextual level, the attention mechanism within the SMILES encoder suggests chemically informative contextual weighting. To assess whether the learned atomic representations reflect chemically relevant patterns rather than spurious statistical correlations, the encoder's attention weights were visualized to identify the substructures emphasized when inferring a given atom's chemical environment. As shown in Fig. 3e, the attention patterns are consistent with known electronic and spatial effects. For *α,β*-unsaturated carbonyl derivatives (Fig. 3e i), the model assigns higher weights to the carbonyl group and the conjugated C=C unit, suggesting sensitivity to structural features that modulate the local shielding environment of the target carbon. This is consistent with the tendency of conjugated carbonyl systems to shift nearby $^{13}C$ resonances downfield. In fused heterocyclic systems (Fig. 3e ii), the attention map further suggests that the model captures the combined influence of aromatic magnetic anisotropy and heteroatom-dependent electronic effects, including those associated with adjacent nitrogen atoms. Furthermore, in a fused N-containing aromatic system (Fig. 3e iii), the model highlights the local heteroaromatic π-framework around the target carbon[53].

Taken together, these analyses suggest that CLAIM does not overcome topology-only limitations by memorizing global correlations, but by using NMR-guided supervision to disentangle chemically distinct local environments that are otherwise merged in conventional topological representations.

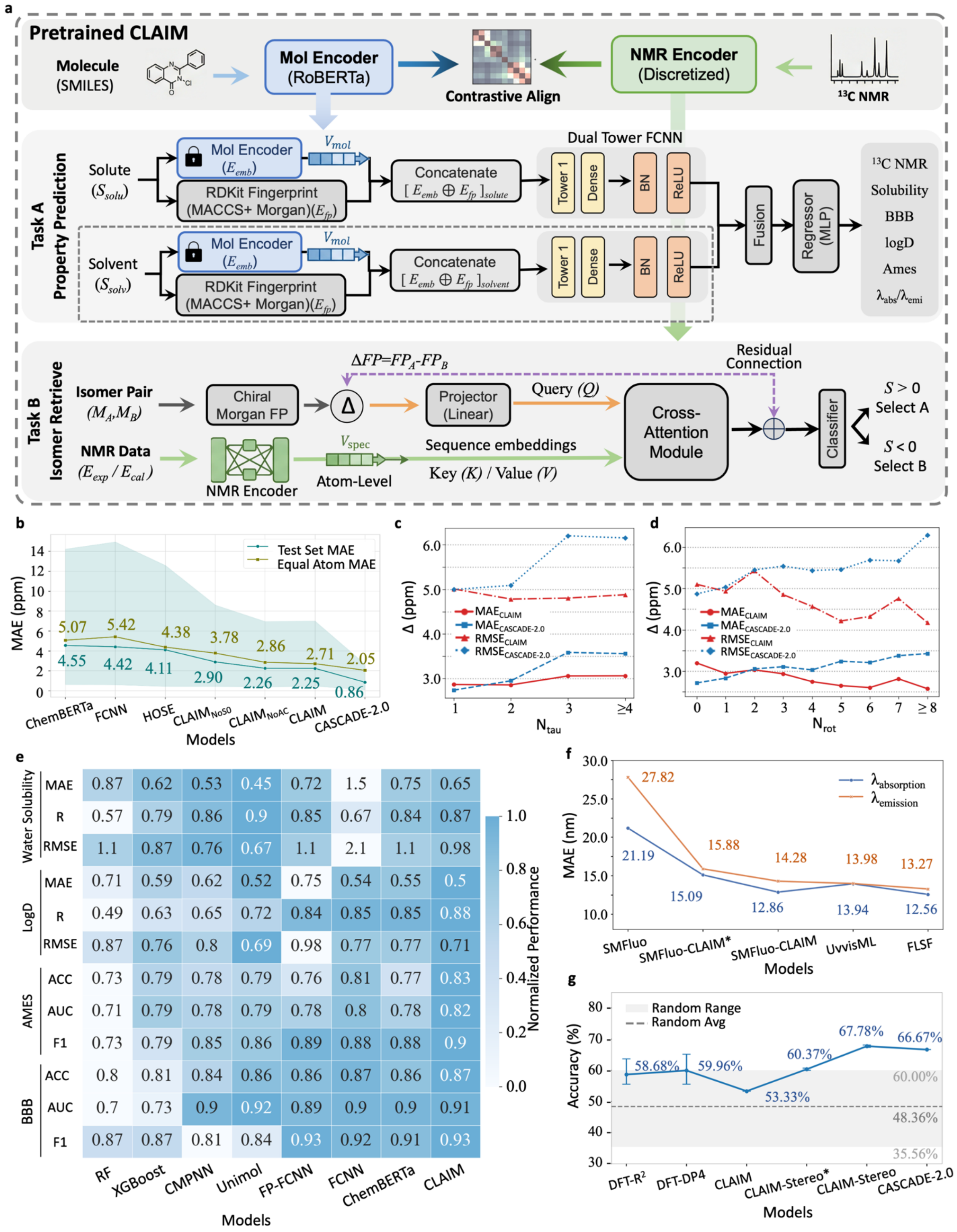

**Figure 4. Evaluation of the CLAIM representation in downstream tasks. a.** Schematic diagram of downstream task workflows. Property Prediction employs a task-adaptive architecture based on frozen CLAIM representations. For fluorescence spectral prediction, a dual-tower FCNN processes solute-solvent pairs by concatenating molecular embeddings with RDKit[54] fingerprints. General molecular properties utilize a single-tower variant with concatenated features. $^{13}C$ NMR chemical shift prediction relies exclusively on CLAIM embeddings processed by a simple MLP, without auxiliary molecular fingerprints. **b.** Benchmarking of $^{13}C$ NMR prediction accuracy. Comparison of MAE values (in ppm) across baseline methods and CLAIM ablation studies. The teal and olive lines correspond to the evaluation results on the Test Set and Equal Atom Set, respectively. The light blue background highlights the error distribution (min-max range) for the Test Set. ChemBERTa uses the ChemBERTa-77M[15] pretrained model as the encoder. $CLAIM_{NoS0}$ and $CLAIM_{NoAC}$ are ablated variants that use CLAIM as the encoder but remove atom-level alignment and atom type/confidence score information, respectively. CLAIM represents the full model with CLAIM as the encoder, whereas FCNN (Fully Connected Neural Network) uses a randomly initialized encoder. **c.** Prediction errors of $^{13}C$ NMR chemical

shifts by the CLAIM-based FCNN and CASCADE-2.0[45] models as a function of the number of isomers ($N_{tau}$). Red lines denote the CLAIM-based model and blue lines denote CASCADE-2.0, with MAE and RMSE represented by solid/dashed and dash-dotted/dotted lines, respectively. **d.** Corresponding prediction errors as a function of the number of rotatable bonds ($N_{rot}$). **e.** Heatmap summarizing the performance of different models on the Water Solubility, LogD, AMES, and BBB datasets[55]. All evaluation metrics were normalized, and inverse normalization was applied to MAE and RMSE so that darker colors uniformly indicate better performance. RF refers to Random Forest. FP-FCNN is a fingerprint-based fully connected neural network. XGBoost, CMPNN[56], and Uni-Mol[57] were used as baseline models. **f.** Comparison of fluorescence property prediction errors. Displays the Mean Absolute Error (MAE) of different models in predicting emission (orange) and absorption (blue) wavelengths. Compared models include: SMFluo[9] (FCNN baseline), SMFluo-CLAIM (incorporating CLAIM representations), SMFluo-CLAIM* (ablation version removing Stage 0 and atom type/confidence representations), as well as UvvisML[17] and FLSF[4]. **g.** Box plot of stereoisomer retrieval accuracy. Comparison baselines include: DFT-$R^2$/DP4[50], CASCADE-2.0[45], CLAIM, CLAIM-Stereo, and CLAIM-Stereo* (randomly initialized NMR encoder). The gray shaded area and dashed line indicate the accuracy range and mean of random guessing.

## Validation in dynamic and stereochemically sensitive settings

To test whether CLAIM restores chemical distinctions that are lost in topology-only representations, we evaluated it in settings where such distinctions are most difficult to preserve: dynamic molecular regimes and stereochemically sensitive discrimination. We first examined near-domain performance on $^{13}C$ NMR chemical-shift prediction, using the benchmark overlapping NMRSDB-Cal and NMRSDB-Exp. CLAIM performed favorably relative to empirical topology-based methods and ChemBERTa[15], while remaining competitive with more computationally intensive structure-based approaches (Fig. 4b). Ablation analysis further showed that both atom-level alignment and atom type/confidence information contribute materially to performance, indicating that the gain arises from physically structured pretraining rather than encoder capacity alone.

This advantage became clearer in molecules containing topologically equivalent yet chemically distinct carbons, which directly probe the representational collapse highlighted above. On this benchmark, conventional 2D methods showed a marked decline in performance, whereas CLAIM remained comparatively robust (Fig. 4b). Although CASCADE-2.0[45] retained an advantage on this restricted set, its errors increased more strongly with structural complexity. When stratified by the number of isomers and the number of rotatable bonds, CASCADE-2.0 showed progressively worsening prediction error with rising tautomeric multiplicity and conformational flexibility, whereas the CLAIM-based model remained comparatively stable across these regimes (Fig. 4c,d). These results are consistent with the distinct inductive biases of the two approaches: explicit 3D conformer-based models remain highly effective when local stereochemical environments can be approximated reliably by enumerated structures, but become less robust when the observed signal increasingly reflects conformational averaging and tautomeric exchange. Consistent with this interpretation, CLAIM also achieved the strongest overall performance in the large-scale benchmark. Together, these results indicate that physically aligned representations are particularly advantageous in dynamic chemical spaces where static 3D descriptions become less reliable.

We next asked whether the same restored chemical resolution could support a more stringent reverse-inference task, namely stereoisomer discrimination. To this end, we constructed CLAIM-Stereo using a delta-learning framework (Fig. 4a, Task B), in which relative spectral perturbations between isomeric pairs were modeled using NMRSDB-Cal together with stochastic SMILES augmentation. In this framework, chiral Morgan fingerprints[58] serve as topological anchors for symmetry breaking, whereas the pretrained NMR encoder provides the physically informed mapping from structural differences to observable spectral perturbations.

Under this reverse-inference setting, CLAIM-Stereo outperformed both the high-precision DFT baseline and CASCADE-2.0 in stereoisomer retrieval accuracy (Fig. 4g). A control model combining chiral fingerprints with a randomly initialized NMR encoder performed substantially worse, indicating that explicit 2D chiral descriptors alone are insufficient for high-fidelity stereochemical discrimination. Cross-attention analysis further showed that CLAIM preferentially focused on stereochemically sensitive regions rather than attending uniformly across the molecule. The resulting attention patterns were consistent with chemical intuition and with DFT-based shift variations among candidate isomers, supporting that the pretrained representation captures long-range stereoelectronic effects relevant to stereochemical discrimination.

Taken together, these validations show that the benefit of NMR-aligned pretraining is most evident where conventional topological representations are prone to fail: in dynamic regimes shaped by conformational averaging and in discrimination tasks governed by subtle stereochemical perturbations. Rather than merely improving near-domain spectral fitting, CLAIM restores chemically meaningful resolution in precisely those settings where topology alone becomes insufficient, providing a basis for transfer beyond direct spectroscopic tasks.

## Transfer beyond spectroscopy

Because CLAIM restores local chemical distinctions rather than merely fitting spectral values, we next examined whether this physically aligned representation transfers to downstream molecular tasks beyond spectroscopy. We first evaluated standard ADMET prediction tasks from PharmaBench[55] (Fig. 4e). A clear contrast emerged between classification and regression settings. In the classification tasks, CLAIM-based model showed the strongest overall performance on both AMES toxicity and blood-brain barrier (BBB) permeability, achieving the best results across all three reported metrics on AMES and the highest accuracy and F1 score on BBB, while remaining competitive in AUC. In the regression tasks, the pattern was more differentiated. For water solubility, CLAIM-based improved substantially over fingerprint-based, ChemBERTa-based and randomly initialized neural baselines, but did not surpass the strongest structure-aware models. By contrast, on LogD, CLAIM delivered the best overall performance among the compared methods, with the lowest prediction error and the strongest correlation. These findings indicate that the information captured through NMR-aligned pretraining is not confined to reconstructing spectroscopic observables, but instead reflects a more transferable description of local chemical environment that supports downstream inference on molecular properties with distinct biological and physicochemical determinants.

We further asked whether this restored chemical resolution extends to an orthogonal property regime, namely fluorescence photophysical prediction. Like NMR, fluorescence absorption and emission are sensitive to electron distribution, conformational relaxation and environment-dependent excited-state effects[59]. Embeddings from the frozen molecular encoder were concatenated with molecular fingerprints and used in a dual-tower Fully Connected Neural Network (FCNN) to predict maximum absorption and emission wavelengths (Fig. 4f), thereby assessing whether CLAIM captures information relevant beyond ground-state NMR observables. Incorporation of CLAIM representations substantially reduced prediction errors relative to the fingerprint-only baseline and brought performance close to specialized graph-based models. This improvement was attenuated in the ablated variant lacking atom-level alignment pretraining, indicating that the gains do not arise solely from increased model capacity, but from the physically structured representation established during pretraining. Together with the ADMET results, these findings show that CLAIM supports forward transfer beyond spectroscopy because the representation learned from NMR alignment captures chemically relevant local environments rather than a task-specific spectral shortcut, with particularly clear advantages in discriminative property prediction and in tasks sensitive to coupled electronic and conformational effects.

## Conclusion

High-confidence NMR curation first made visible a recurrent representational failure that had been obscured by noisy and unassigned data: atoms that are equivalent in 2D topology can remain experimentally distinct in their real chemical environments. This observation clarifies a central limitation of conventional molecular representations. Efficient topological encodings often collapse chemically distinct local environments, whereas explicit 3D approaches improve structural specificity but remain constrained by computational cost and by their dependence on static geometric states. Together, these results define the bottleneck addressed in this study as a trade-off between structural specificity and solution-state realism.

CLAIM was introduced as a physically aligned route beyond this bottleneck. Instead of increasing geometric fidelity through explicit conformer generation, the framework uses atom-resolved $^{13}$C NMR observables to supervise representation learning from efficient topological inputs. In this setting, NMR does not simply provide an auxiliary modality; it enables the model to disentangle chemically distinct local environments that remain merged under topology-only encoding, thereby recovering chemical resolution in latent space. The retrieval and ablation results indicate that this gain arises primarily from atom-level physical supervision rather than from encoder capacity alone.

Validation across near-domain prediction, stereochemically sensitive discrimination and downstream transfer further supports this interpretation. In $^{13}$C NMR prediction, CLAIM remained particularly robust in flexible and tautomeric systems, where static geometry-based models became less reliable. In stereoisomer discrimination, the pretrained representation improved retrieval beyond high-precision DFT baselines when combined with explicit chiral anchors, indicating that the learned features capture stereoelectronic distinctions relevant to reverse inference. In ADMET and fluorescence tasks, the same representation transferred beyond direct spectral modelling, suggesting that the information learned from NMR alignment reflects a broader description of local chemical environment rather than a task-specific spectral shortcut. Taken together, these results place CLAIM as a practical bridge between experimentally grounded molecular observation and transferable molecular representation.

At the same time, the present framework also defines its own boundary. Decoupled $^{13}$C NMR provides an effective constraint on chemically distinct local environments and on their time-averaged expression in solution, but it remains less informative for higher-order magnetic equivalence and coupling-sensitive stereochemical relationships. This limitation is reflected most clearly in the reverse-inference setting, where stereochemical discrimination is strengthened by explicit chiral structural anchors. A fuller description of molecular electronic environments will therefore require broader spectroscopic supervision, particularly from $^{1}$H NMR and related observables that report more directly on spin coupling and dynamic exchange. Extending physically aligned representation learning toward such higher-dimensional spectral constraints may further reduce the remaining topological degeneracy and move molecular representation closer to experimentally observed chemical reality.

## Methods

### Adaptive NMR Validator

The Adaptive NMR Validator process begins with atom typing using a predefined set of SMARTS patterns derived from chemical intuition, mapping each carbon atom to a specific chemical environment. For each atom type $T$, the distribution of experimental chemical shifts $X$ is modeled using a Gaussian Mixture Model (GMM). The probability density function is defined as $P(x|T) = \sum_{k=1}^{K} w_k \, \mathcal{N}(x; \mu_k, \sigma_k^2)$, where the number of components $K$ (up to 4) is determined automatically via Bayesian Information Criterion (BIC) minimization. To prevent overfitting to sparse data clusters, Gaussian components with means closer than a threshold (default 5.0 ppm) are iteratively merged, recalculating the pooled variance and weights to maintain statistical integrity. The baseline fitting was performed on pairs of atoms assigned to the same predefined chemical environment across independent experimental records, such that the resulting distribution captured both systematic variation and residual structure-dependent heterogeneity. The GMM decomposition was used only to estimate the systematic-error boundary for downstream validation, rather than to assign mechanistic meaning to each Gaussian component.

Crucially, the validator employs an auto-tuning mechanism to handle the long-tail distributions characteristic of heterogeneous experimental conditions. We introduce a widening factor $\lambda$ that scales the standard deviation of the fitted components ($\sigma'_k = \lambda\sigma_k$). This factor is optimized iteratively for each atom type to ensure that a target proportion (99.9%) of the training data falls within a defined probability threshold, thereby establishing a robust physical boundary. In retrospective database screening, atoms falling below the 0.01 quantile of the tuned confidence distribution for their assigned environment were flagged for manual inspection. This low-confidence subset was used to identify recurrent data issues, including index misalignment and structural misannotation. For a given query molecule with observed shift $x_{obs}$ at atom $i$, the validator calculates a relative likelihood score based on the tuned PDF. To account for finite sampling, a p-value derived is also computed from the cumulative distribution function. A molecule is accepted only if the lowest p-value among all its atoms exceeds a significance level $\alpha$ (0.05), or if the aggregate error probability, calculated via a harmonic mean of individual atomic scores, remains below a tolerance threshold. This ensures that the validator rejects unphysical outliers while accommodating legitimate variations within the chemical space.

## Hybrid NMR Predictor and Data Curation

The Hybrid NMR Predictor synergizes the structural specificity of Hierarchical Organization of Spherical Environments (HOSE)[46] codes with the generalization capability of substructure-based Gaussian Mixture Models (GMMs). The prediction algorithm utilizes a cascade retrieval strategy. For a target atom, statistical moments (mean $\mu$, standard deviation $\sigma$, and count $N$) are retrieved from a hierarchical lookup table indexed by HOSE codes. The search initiates at a maximum spherical radius ($r_{max} = 6$) and iteratively relaxes to a minimum radius ($r_{min} = 3$). Matches are accepted only if the historical sample abundance exceeds a defined threshold ($N \geq 3$). This rule implements the conservative degradation strategy used in curation: high-resolution HOSE environments are preferred whenever sufficient historical support is available, and degradation to lower-resolution shells is permitted only when the abundance criterion is met. Upon failure of the HOSE-based search, the predictor reverts to the SMARTS-based GMM distribution. The predicted uncertainty is modeled using a Root Sum Square (RSS) formulation. The effective standard deviation $\sigma_{eff}$ is calculated as:

$$\sigma_{eff} = \sqrt{\sigma_{pred}^2 + \sigma_{sys}^2} \cdot \lambda_{widen}$$

where $\sigma_{pred}$ is the variance from the chemical environment model, and $\sigma_{sys}$ represents the irreducible systematic variance derived from solvent effects and topological equivalency fluctuations (calibrated to $\approx$ 2.3 ppm). This formulation generates a continuous probability density function for every atom, providing the probabilistic basis for spectral assignment.

The data curation pipeline resolves ambiguities in blind spectral datasets through symmetry-based imputation and algorithmic realignment[60]. First, topological symmetry is handled by grouping carbon atoms via canonical ranking (tie-breaking disabled); experimental chemical shifts are broadcast across topologically equivalent atoms to ensure label completeness. To detect index permutations, the rearrangement inequality is leveraged: a record is flagged as misaligned if sorting the predicted and experimental vectors independently reduces the Mean Absolute Error (MAE) by more than 2.0 ppm relative to the archival alignment.

For automated annotation, predictor selection was based not only on scalar regression accuracy but also on its behavior under rank-based assignment with rejection. In this setting, candidate assignments are accepted only when the predicted ordering provides sufficient separation between adjacent peaks; otherwise, the record is discarded. Explicit 3D or geometry-based predictors were therefore not used for primary curation, because rigid conformational snapshots can introduce spurious shift differences between atoms that are chemically equivalent in solution. Such false distinguishability may improve nominal regression resolution while generating artificial ranking cues that increase the risk of incorrect assignment. By contrast, the prior-guided topology-based predictor preserves true chemical equivalence and is therefore better suited to high-precision annotation under the rejection-based matching scheme.

Identified misaligned records undergo rank-order correction, mapping experimental peaks to atoms based on the rank of their predicted shifts. This reassignment is strictly conditional on spectral resolution; to prevent errors in congested regions, correction is aborted unless the separation between all adjacent predicted peaks exceeds an ambiguity threshold:

$$\min_k \left( y_{(k+1)}^{pred} - y_{(k)}^{pred} \right) \geq 8.0 \, ppm$$

This rejection rule was designed to favor precision over coverage: records were reassigned only when the predictor induced a sufficiently separated ordering, thereby minimizing erroneous correction in spectrally congested regions. Assignments satisfying this resolution constraint are remapped according to the predicted rank order. Finally, the corrected data are validated against the Hybrid NMR Predictor using a distribution widening factor of $\lambda = 1.5$, ensuring only high-confidence assignments are integrated into the training corpus.

## AutoStereoQ Automated Computational Workflow

The AutoStereoQ workflow initiates with the conversion of 2D molecular graphs into initial 3D cartesian coordinates using OpenBabel. To ensure exhaustive coverage of the potential energy surface across varying degrees of molecular flexibility, we implemented four distinct conformational search protocols: GMX, GMX*[61], xTB and RDKit[54]. These 4 protocols were first benchmarked on a small set of 11 structurally diverse natural products to compare conformational coverage and downstream NMR fidelity, and were subsequently used as interchangeable front ends within AutoStereoQ for large-scale evaluation.

The xTB protocol executes molecular dynamics simulations coupled with on-the-fly geometry optimization directly at the GFN0-xTB semi-empirical level *via* the xTB program. The GMX protocol employs GROMACS for trajectory generation and uses GROMACS' built-in energy evaluation for scoring. Similarly, the GMX* protocol uses GROMACS[61] to perform high-temperature molecular dynamics (MD) simulations for topological sampling; extracted snapshots are subsequently evaluated using GFN0-xTB energies (GMX*) to improve electronic accuracy.

Uniquely, the RDKit protocol employs an in-house adaptive systematic search algorithm designed to exhaustively sample local minima. This module identifies non-cyclic rotatable bonds and calculates the theoretical conformer count based on a torsional grid. To balance coverage with computational feasibility, the sampling resolution is dynamically adjusted—starting from $\theta = 30°$ down to a minimum of $10°$—until the total ensemble size falls within a predefined limit ($\mathrm{N} \leq 1000$). Generated conformers are constructed via the cartesian product of dihedral angles and immediately relaxed using the MMFF94 force field. This adaptive torsional grid was introduced to increase orthogonal coverage of low-energy conformational space while keeping the total ensemble size computationally tractable.

Following the initial sampling, conformers from all streams are pooled and subjected to a multi-stage refinement process. The ensemble undergoes iterative clustering and semi-empirical optimization at the GFN0-xTB and GFN2-xTB[62] levels, where xTB[63] serves as a robust and efficient quantum mechanical method for geometry pre-optimization. Redundant structures are pruned based on geometric root-mean-square deviation (RMSD) and energy thresholds. Surviving conformers undergo high-level Density Functional Theory (DFT) geometry optimization using either ORCA 6.0[64] or Gaussian 16 packages[65]. We employed a composite strategy where geometry optimization is performed at a B3LYP/def2TZVP level, followed by single-point NMR shielding tensor calculations using the GIAO (Gauge-Independent Atomic Orbital)[66] method with different levels. To simulate solution-state physics, an implicit solvation model matching the experimental solvent is applied throughout the process. Finally, the chemical shifts are derived by Boltzmann-weighting the shielding tensors of all conformers within a chemically relevant energy window (5.0 kcal/mol) at 298.15 K. The resulting benchmark was used to compare cost–accuracy trade-offs across 18 computational strategy combinations and to evaluate the extent to which ensemble-based explicit 3D workflows improve regression fidelity and stereochemical discrimination. This automated pipeline ensures reproducibility and allows for the systematic generation of the NMRSDB-Cal database, providing a theoretical ground truth for examining the limitations of static geometric representations.

## Training Data Processing and Augmentation

Stochastic SMILES Enumeration was implemented using the RDKit framework to process molecular inputs. Atom indices within the molecular graph were randomly renumbered to generate non-canonical SMILES variants, with N=5 variants produced for the training set and *N=2* for the validation set per molecule. To map the sequential tokens to specific physical atoms, the standard InChI atom layer ranking was computed for each molecular graph, establishing the specific atomic indexing used for feature alignment.

Spectroscopic features were processed via statistical noise injection and the assignment of chemical priors. Gaussian noise $N(\mu, \sigma^2)$ was superimposed onto the raw $^{13}$C NMR chemical shift values, utilizing parameters μ=0.938 and σ=1.34 derived from error distribution analysis. For each carbon atom, a discrete atom type was assigned from 83 predefined classes, along with a continuous confidence score (normalized between 0 and 1). These values were formatted as paired attributes to accompany the shift data.

The final dataset was derived from the NMRSDB-Exp database and partitioned into training, validation, and testing sets according to a 7:2:1 ratio, adhering to a structural non-overlap principle. The data were organized into dual formats: an atom-level view containing SMILES sub-sequences paired with specific atomic shifts, and a molecule-level view containing full SMILES sequences paired with complete spectra. The resulting dataset comprises 648,121 molecule-spectrum pairs for training, 29,685 for validation, and 14,841 for testing.

## CLAIM Model Architecture and Two-Stage Training

The CLAIM (Contrastive Learning for Atom-to-molecule Inference of Molecular NMR) model adopts a dual-tower architecture comprising a molecule encoder and an NMR encoder. The molecule encoder is trained by fine-tuning the pre-trained transformer backbone (ChemBERTa[67]) to process SMILES sequences. Atom-specific features corresponding to the NMR signals are extracted using a gathering mechanism that maps the tokenized sub-word embeddings back to their respective carbon-atom indices based on InChI-rank alignment. This ensures that the learned representations correspond directly to physical atomic centers.

The NMR encoder is designed to handle the specific nature of spectroscopic data. Instead of treating chemical shifts as continuous floating-point values, we employed a high-resolution grid encoder. The spectral range (typically -50 to 350 ppm) is discretized into 4000 bins, and each shift is mapped to a learnable high-dimensional embedding. Furthermore, we integrated the hierarchical chemical priors by adding learnable atom type embeddings (derived from the 83 classes defined in the validator) to the shift embeddings. Crucially, the confidence score obtained from the Adaptive NMR Validator is injected as a multiplicative gating factor into the feature space ($h_{nmr} = h_{feat} \times c_{score}$), allowing the model to dynamically down-weight noisy or uncertain experimental data during training.

CLAIM is trained *via* a two-stage curriculum. Stage 0 (atom-level alignment) focuses on the microscopic physics. We construct a contrastive loss between the gathered atom embeddings from the molecule encoder and the corresponding encoded features from the NMR encoder. Stage 1 (global-level alignment) extends this to the molecular scale. We employ a global aggregator module utilizing multi-head attention pooling to synthesize a holistic molecule vector ($V_{mol}$) and a spectrum vector ($V_{spec}$) from their respective atomic components. A symmetric InfoNCE loss is applied to maximize the similarity between matched molecule-spectrum pairs while minimizing it for mismatched pairs within the batch.

## Downstream Property Prediction

For downstream property prediction tasks, specific regression architectures are constructed on top of the frozen, pre-trained CLAIM molecular encoder. For $^{13}$C NMR prediction, 768-dimensional atom-level embeddings are extracted from the molecular backbone using sequence gather indices. These embeddings are directly fed into a pyramid Multi-Layer Perceptron (MLP) head comprising two hidden layers with dimensions of 512, and 256, respectively. Each hidden layer sequentially applies a linear transformation, Layer Normalization, Gaussian Error Linear Unit (GELU) activation, and Dropout (0.1), terminating in a final linear layer for continuous chemical shift regression. For the prediction of the four macroscopic physicochemical properties, a hybrid feature representation is constructed by concatenating the 768-dimensional molecular-level [CLS] token from the encoder with explicit RDKit molecular fingerprints (MACCS and Morgan, totaling 200 bits). This concatenated high-dimensional vector is processed through a Fully Connected Neural Network (FCNN) consisting of multiple dense layers (512 units) equipped with Batch Normalization, LeakyReLU activation, and Dropout (0.2), followed by a final regression head.

The fluorescence property prediction adopts an identical foundational architecture but is expanded into a dual-tower FCNN to process solute and solvent features in parallel. The hybrid representations for both the solute and the solvent are fed into their respective dense branches, and the resulting processed vectors are concatenated and passed through a fusion MLP prior to the regression head. The fluorescence regression model employs a LogCosh loss function, and all downstream networks are optimized using the Adam optimizer coupled with a cosine decay learning rate schedule.

### Delta Learning for Stereochemical Discrimination

CLAIM-Stereo adapts the input modality to explicit chiral morgan fingerprints (radius 2, 1024 bits) while leveraging the pre-trained NMR encoder as a physics-informed feature extractor. The model adopts a delta learning paradigm to learn the relative spectral perturbations induced by stereochemical inversion. Specifically, for a pair of candidate isomers $A$ and $B$, the fingerprint difference vector is computed as $\Delta FP = FP_A - FP_B$. This differential feature is projected into a latent space *via* a Multi-Layer Perceptron (MLP) and serves as the Query in a cross-attention module. The Keys and Values are derived from the NMR peak sequence features extracted by the pre-trained NMR encoder. The network is trained using a Margin Ranking Loss with a margin of $\alpha = 0.5$. For a given query spectrum $S$ and a pair of candidate structures where structure $A$ is the ground truth, the loss is defined as:

$$L = \max\big(0, -y \cdot \big(f(S, A, B)\big) + \alpha\big)$$

where $y = 1$ if $A$ is the positive sample, and $f(S, A, B)$ is the scalar score difference predicted by the model. For a query spectrum $S$ derived from an anchor molecule, positive samples are defined as structures possessing the identical relative stereoconfiguration or its exact enantiomer, as they inherently yield indistinguishable NMR signals in achiral environments. Negative samples are systematically selected from the remaining diastereomers within the same stereoisomeric group, which exhibit distinct and detectable spectral characteristics. To preserve the physical manifold learned during pre-training, a conservative fine-tuning strategy with a low learning rate ($10^{-4}$) and a warm-up ratio of 0.3 is employed. Data augmentation is applied by randomly adding noise to the NMR shifts ($k \in [0.9, 1.1], b \in [-2.0, 2.0]$) to simulate experimental variance and force the model to learn robust relative features.

## Data availability

All the data and codes used in this study are available at XXX.

## Acknowledgments

This work was supported by the Pioneer and Leading Goose R&D Program of Zhejiang (2025C02087), the National Natural Science Foundation of China (82473798), the Key Research and Development Program of Hainan (ZDYF2025SHFZ053), the China Postdoctoral Science Foundation (2025M783584), the Seed Fund Cultivation Project of Ocean College, Zhejiang University (2025YQ005), the Scientific Research Fund of Zhejiang University (XY2025062), and the Zhejiang Provincial Department of Marine Economic Development Project (2026HYA10001). Computations were supported by the High-performance Computing Platform of YZBSTCACC, the Information Technology Center, Zhejiang University, and China Mobile Zhejiang Co., Ltd., Hangzhou Branch.

## Author contributions

J.F. conceived and designed the algorithm, conducted the evaluation and analyses, and wrote the initial manuscript draft, with figures preparation. Z.Y. performed part of the data processing. C.M. and Y.J. carried out data annotation and assisted with data analysis. L.M. provided computational resources. Y.H. conducted chemical analysis and data interpretation, and performed data validation. W.D. edited and refined the manuscript. Z.M. designed and supervised the project. All authors reviewed and edited the manuscript and approved the final version.

## Competing interests

The authors declare no competing interests.

## Additional information

**Supplementary information** The online version contains supplementary material available at XXX.

Correspondence and requests for materials should be addressed to XXX.

TOC

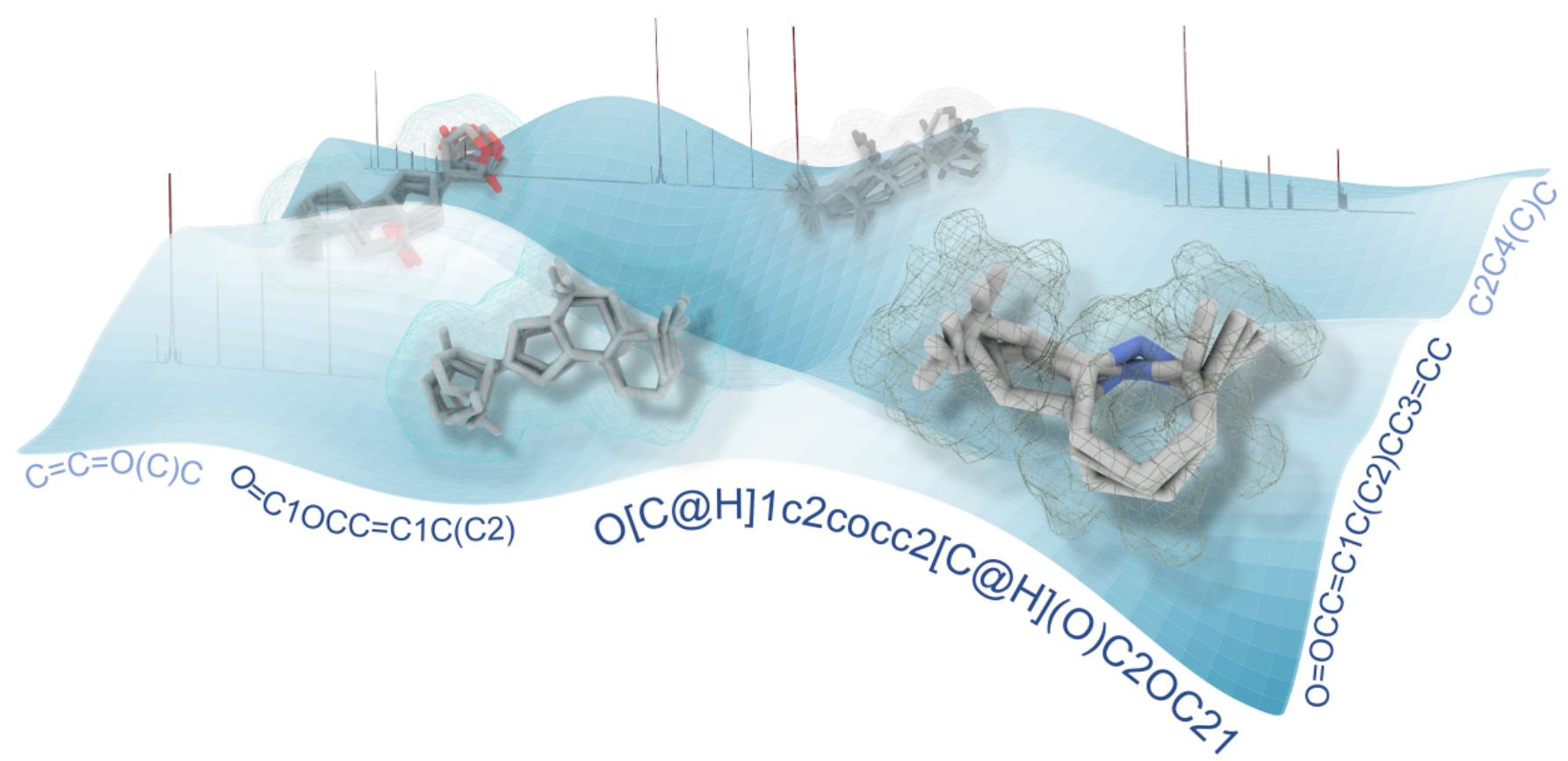

C=C=O(C)C
O=C1OCC=C1C(C2)
O[C@H]1c2cocc2[C@H](O)C2OC21
O=OCC=C1C(C2)CC3=CC
C2C4(C)C